\documentclass[12pt]{article}
\usepackage{epsfig}
\usepackage{lineno}
\usepackage{cite}
\usepackage{rotating}
\usepackage{booktabs}
\usepackage{lscape}
\usepackage{amsmath}
\usepackage{multirow}
\usepackage[usenames]{color}

\usepackage{color}

\usepackage{cite}

\begin{document}

\begin{center}

\vspace*{1.0cm}

{\Large \bf{Measurements of ZnWO$_4$ anisotropic response to nuclear recoils for the ADAMO project}}

\vspace*{1.0cm}

{\bf P.~Belli$^{a,b}$, R.~Bernabei$^{a,b,}$\footnote{Corresponding
author. {\it E-mail address:} rita.bernabei@roma2.infn.it.},
F.~Cappella$^{c,d}$,\\
V.~Caracciolo$^{a,b,e}$, R.~Cerulli$^{a,b}$, N.~Cherubini$^f$, F.~A.~Danevich$^g$,\\
A.~Incicchitti$^{c,d}$, D.~V.~Kasperovych$^g$, \\ 
V.~Merlo$^{a,b}$, E.~Piccinelli$^f$, O.~G.~Polischuk$^{g}$ and 
V.~I.~Tretyak$^g$}

\end{center}

\vskip 0.3cm

{\footnotesize
\noindent $^a$ INFN sezione Roma ``Tor Vergata'', I-00133 Rome, Italy\\
$^b$ Dipartimento di Fisica, Universit\`a di Roma ``Tor Vergata'', 
I-00133, Rome, Italy\\
$^c$ INFN sezione Roma, I-00185 Rome, Italy\\
$^d$ Dipartimento di Fisica, Universit\`a di Roma ``La Sapienza'', I-00185 
Rome, Italy\\
$^e$ INFN, Laboratori Nazionali del Gran Sasso, I-67100 Assergi (AQ), 
Italy\\
$^f$ Enea, Italian National Agency for New Technologies, Energy and Sustainable Economic \\
\indent   Development, C.R: Casaccia, Roma 00123, Italy\\
$^g$ Institute for Nuclear Research of NASU, 03028 Kyiv, Ukraine\\

\vskip 0.5cm

\begin{abstract}
Anisotropic scintillators can offer a unique possibility to exploit
the so-called directionality approach in order to investigate the presence of those Dark Matter (DM) candidates 
inducing nuclear recoils. In fact, their use can overcome the difficulty of detecting extremely short nuclear recoil traces. 
In this paper we present recent measurements performed on the anisotropic response of a ZnWO$_4$ crystal scintillator 
to nuclear recoils, in the framework of the ADAMO project.
The anisotropic features of the ZnWO$_4$ crystal scintillators
were initially measured with $\alpha$ particles; those results have been also confirmed by the additional measurements 
presented here.
The experimental nuclear recoil data were obtained by using a neutron generator at ENEA-CASACCIA and neutron detectors 
to tag the scattered neutrons; in particular, the quenching factor
values for nuclear recoils along different crystallographic axes have been determined for three different neutron scattering angles
(i.e. nuclear recoils energies).
From these measurements, the anisotropy of the light response for nuclear recoils in the ZnWO$_4$ crystal scintillator has been determined at 5.4 standard deviations.
\end{abstract}

\vskip 0.4cm  


\normalsize

\section{Introduction}

Astrophysical observations have pointed out the presence of Dark Matter (DM) on all astrophysical scales and many 
arguments have been suggested that a large fraction of it should be in form of relic DM particles. In the direct search for 
DM, a positive model independent result has been obtained with high confidence level by DAMA/NaI, DAMA/LIBRA-phase1 and 
DAMA/LI\-BRA-phase2 \cite{RNC,modlibra,modlibra2,modlibra3,npae18} exploiting the DM annual modulation signature with 
ultra-low-background NaI(Tl) target. 
This model independent evidence is compatible with a wide set of scenarios regarding the nature of 
the DM candidates and related astrophysical, nuclear and particle physics (see e.g. 
Ref. \cite{modep19} and references therein; also numerous references on DM candidates and scenarios exist).
It is worth noting that the used approach and target assure sensitivity both to DM candidate particles inducing 
nuclear recoils, as well as to those giving rise to electromagnetic radiation.

Another possible approach to DM investigation is the directionality \cite{Sperg88}. In principle, it is effective for those DM candidate 
particles 
able to induce nuclear recoils, and thus potentially giving complementary information. 
This approach studies the correlation between the arrival direction of the DM particles, 
through the induced nuclear recoils, and the Earth motion in the galactic rest frame. In fact, the dynamics of the rotation of 
the Milky Way galactic disc through the halo of DM causes the Earth to experience a wind of DM particles apparently flowing 
along a direction opposite to that of the solar motion relative to the DM halo. However, because of the Earth's rotation around 
its axis, their average direction with respect to an observer fixed on the Earth changes during the sidereal day. 
Nuclear recoils are expected to be strongly correlated with the particle impinging direction, and thus 
the study of the nuclear recoils direction can offer a tool to point out the presence 
of the considered type of DM candidate particles. However, the range of recoiling nuclei is of the order of a few mm  in low 
pressure Time Projection Chambers  and typically of order of $\mu$m  in 
solid detectors. These practical limitations, affecting possible experiments aiming at measuring recoil tracks, can be overcome 
by using anisotropic scintillation detectors in a suitable low background experimental set-up located deep underground. In 
fact, in this case there is no need of a track detection and recognition, since the information on the presence of the 
considered DM 
candidate particles is given by the variation of the measured counting rate (for details see e.g. in Refs. \cite{direz1,direz2,direz3}) 
during the 
sidereal day when the light 
output (as well as the pulse shape) vary depending on the direction of the impinging DM particles with respect to the crystal 
axes.  Moreover, the use of inorganic crystal scintillators takes potentially advantage in terms of high duty cycle, high 
stability, etc.

The use of anisotropic scintillators to study the directionality signature was proposed for the first time in Ref. \cite{direz1} 
and 
revisited in \cite{direz2}, where the case of anthracene detector 
was preliminarily analysed and several practical difficulties 
in the 
feasibility of such an experiment were underlined. Nevertheless, the authors suggested that competitive sensitivities could be 
reached with new devoted R\&D's for the development of anisotropic scintillators having significantly larger size, higher 
light response, better stiffness, higher atomic weights and anisotropic features similar as -- or better than -- those of the 
anthracene scintillator. 

A possibility to use anisotropic properties of ZnWO$_4$ scintillator to search for diurnal variation of the DM 
counting rate was pointed out for the first time in Ref. \cite{Danev05}.
More re\-cently, mea\-sure\-ments and R\&Ds have shown that the ZnWO$_4$ can be a scintillator offering suitable features 
\cite{direz3,zwo1,zwo2,zwo3,zwo4} for the purpose. In particular, the light output due to heavy particles (p, $\alpha$, nuclear 
recoils) depends on the direction of 
such particles with respect to the crystal axes.
Such an anisotropic effect has been ascribed to preferred directions of the 
excitons' propagation in the crystal lattice affecting the dynamic of the scintillation 
mechanism \cite{Birks94,Heck59,Heck61,Kien61,Tsuk62,Tsuk66,Krat70,Broo74}.
In particular, the presence of heavy ionizing particles with a preferred direction (like recoil nuclei induced by the 
considered type of DM candidates) could be discriminated from the electromagnetic background by comparing the low energy 
distributions measured using different orientations of the crystal axes along the day \cite{direz1,direz2,direz3}. 

In the light of this, the ADAMO (Anisotropic detectors for DArk Matter Observation) project \cite{direz3,zwoconf8} was considered 
and R\&Ds have been progressed. 
In particular, here the final results on the measurements performed with a neutron generator at ENEA-CASACCIA 
and neutron detectors to tag the scattered neutrons are presented; 
these results allow to quantify the ZnWO$_4$ anisotropy. Preliminary results in various conditions have been
previously presented \cite{zwoconf1,zwoconf2,zwoconf3,zwo4,zwoconf5,zwo3,zwoconf7,zwoconf8,zwoconf9}.

\section{The ZnWO$_4$ crystal scintillator}

The used ZnWO$_4$ crystal scintillator has a size of ($10\times10\times10.4$) mm$^3$ and a mass of 7.99 g.
The small size of the crystal is necessary to reduce as much as possible the multiple neutron scattering events
in the measurements with the neutron generator.

The crystal has been produced in the framework of the ADAMO project for realizing high level radio-purity ZnWO$_4$ crystal scintillators,
with reasonably high optical and scintillation properties. 
In particular, it has been obtained by second crystallization (using low-thermal gradient Czochralski technique \cite{Shl2017}) 
from zinc tungstate crystals made from tungsten oxide 
additionally purified 
by double sublimation of tungsten chlorides. Only top parts of first crystallization crystal boules were used to produce the 
second crystallization crystal boule. 
The boule then has been cut in various parts among which there was the scintillator used here.
The crystal sample was etched in a solution of 80\% of HNO$_3$  (69.5\%) and 20\% of HF (40\%)
at temperature 85 $^\circ$C over 4 hours\footnote{The etching procedure was applied to remove possible surface radioactive contamination of 3 samples of ZnWO$_4$ scintillators, including the one used in the present study. Two other samples of larger mass, produced from the same crystal boule, were installed in a low-background set-up to investigate radioactive contamination of the material. The measurements are similar to those described in Ref. \cite{zwo4}.}.
The crystallographic axes were identified by the producer and experimentally verified.

\vspace{0.2cm}

The ZnWO$_4$ crystal has been coupled in case of neutron ($\alpha$) measurements to two (one)
high-speed response HAMAMATSU H11934-200 photomultipliers (PMTs), each one having squared window ($30 \times 
30$) mm$^2$, effective area ($23 \times 23$) mm$^2$ and Ultra Bialkali photocathode.
The quantum efficiency is equal to about 43\% at the peak sensitivity (400 nm) and to 
around 25\% at the 500 nm (close to the scintillation emission spectrum of the ZnWO$_4$).

The crystal and one PMT are shown in Fig. \ref{fg:crystal}. It should be noted the optical transparency of the realized ZnWO$_4$ crystal sample, which 
is a very important issue to exploit such a kind of crystal in the considered framework.

\begin{figure}[!th]
	\begin{center}
		\includegraphics [width=0.6\textwidth]{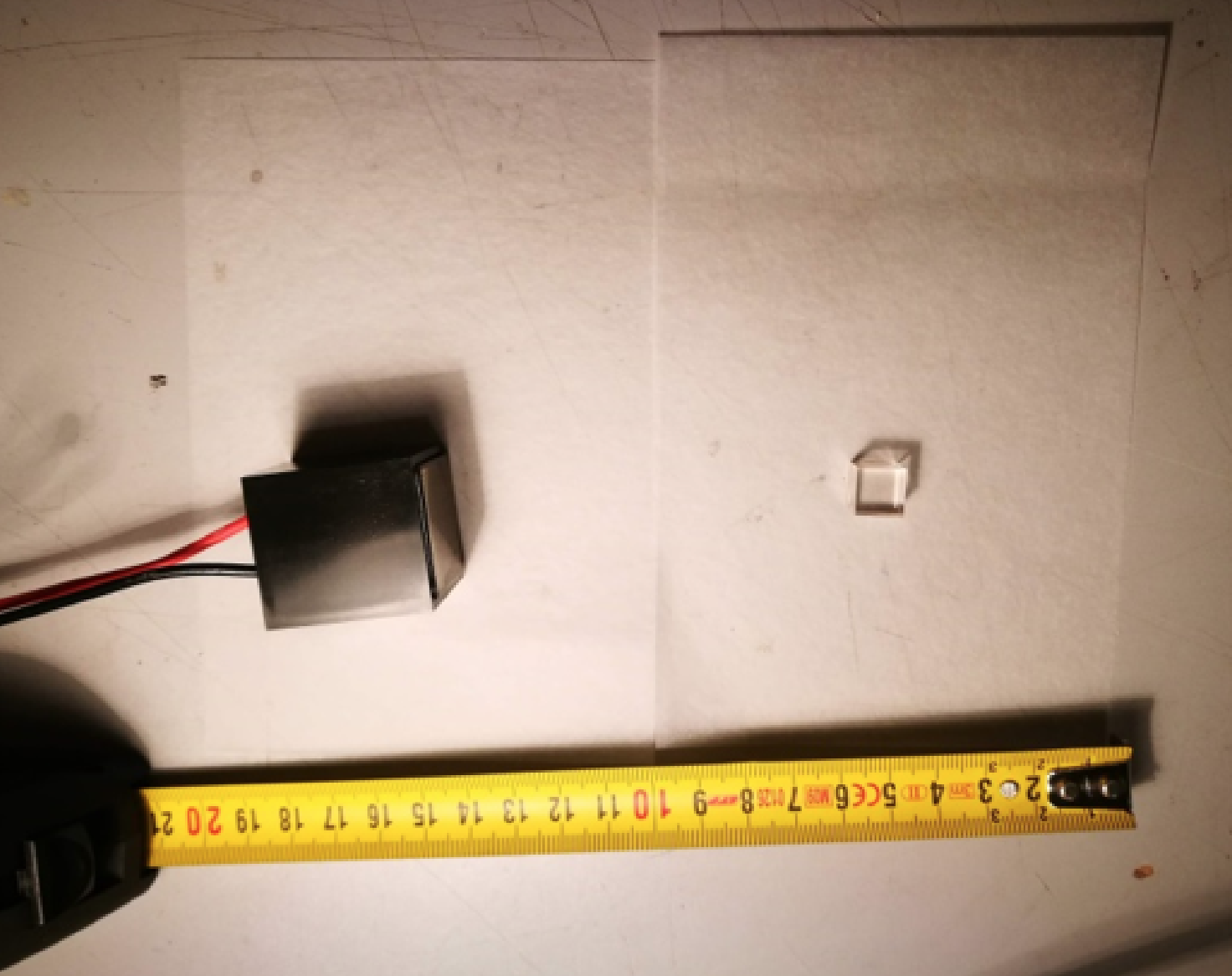}
	\end{center}
	\vspace{-0.8cm}
	\caption{\footnotesize{Picture of the ZnWO$_4$ crystal and of one of the two used HAMAMATSU 
H11934-200 PMTs.
	}}
	\label{fg:crystal}   
\end{figure}

\section{Measurements with $\alpha$ particles}

In Ref. \cite{Danev05} pioneer studies on the anisotropic features of a ZnWO$_4$ crystal have been performed showing that the light 
response and the pulse shape of such a scintillator depend on the impinging direction
of the $\alpha$ particles with respect to the crystal axes. In the present work new dedicated measurements with $\alpha$ particles on the 
ZnWO$_4$ detector, used in the measurements with neutron 
generator (discussed in Sect. \ref{meas_n}), have been performed and compared 
with those of Ref. \cite{Danev05,direz3}.
 
In these measurements the small ZnWO$_4$ crystal, described above, has been coupled to a HAMAMATSU H11934-200 PMT by optical 
grease, wrapped by teflon tape and installed in a ``black box''.

The signal from the PMT has been acquired using a LeCroy WaveSurf24X-sA oscilloscope (4 channels, 2.5 GSamples/s, 200 MHz) and
recorded in a time window of 100 $\mu s$; an event-by-event data acquisition system stored the
pulse profiles of the events.

The measurements have been performed in the MeV region with the help of a collimated 
beam of $\alpha$ particles from an $^{241}$Am source and using various 
sets of thin mylar films as absorbers. The latter ones were placed perpendicular to the $\alpha$ beam 
(fixed by a specific slot into the collimator, 
which was developed for this purpose). The energy scale of each measurement has been calibrated by
using $^{137}$Cs and $^{22}$Na $\gamma$ quanta.

The energies of $\alpha$ particles, considering also the absorber mylar films previously mentioned, were determined with the help 
of a dedicated Silicon detector working in vacuum (CANBERRA Alpha Spectrometer model 7401VR).

Typical energy distributions of the $\alpha$ particles impinging along the three axes of the crystal 
are shown in Fig. \ref{fg:a_peaks}.
The ZnWO$_4$ crystal was irradiated in the directions perpendicular to the  (100), (001) and (010) crystal planes:
hereafter crystal axes I (blue on-line), II (green on-line) and III (red on-line), respectively in Fig. \ref{fg:a_peaks}.

\begin{figure}[!ht]
  \centering
   \vspace{-0.4cm}
  \includegraphics[width=0.7\textwidth]{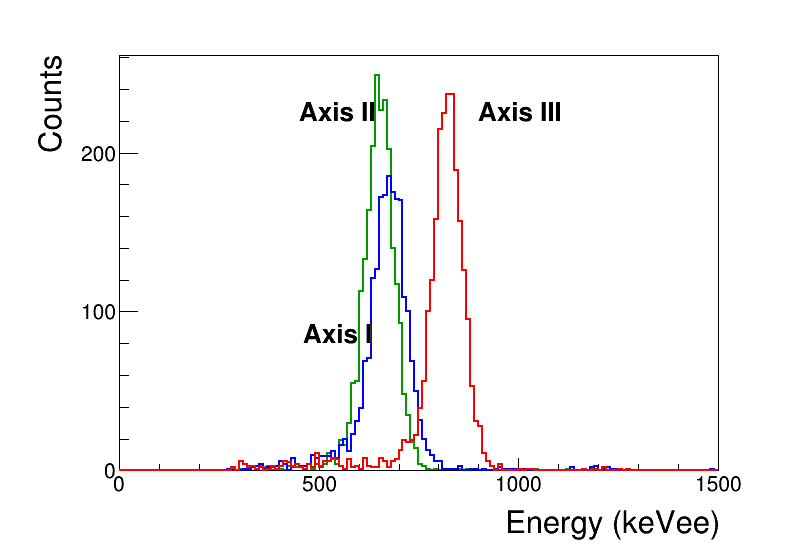}
   \vspace{-0.4cm}
   \caption{\footnotesize{Energy spectra of 4.63 MeV $\alpha$ particles impinging along the three axes of the crystal.
The crystal was irradiated in the directions along the crystal axes I (blue on-line), II (green on-line), and III (red on-line), respectively. 
}}
\label{fg:a_peaks}
\end{figure}

Figure \ref{fg:q_alpha} shows the dependence of the quenching factor Q.F.\footnote{The quenching factor Q.F. describes the response of a scintillator to heavy ionizing particles;
in details, it is the ratio between the detected energy in the energy scale measured 
with $\gamma$ sources to the energy of the heavy ionizing particle.
} for $\alpha$ particles\footnote{Many times it is also named $\alpha/\beta$ ratio.} on the energy and on the direction of the $\alpha$ beam
relatively to the crystal planes in the ZnWO$_4$ crystal.

\begin{figure}[!ht]
\centering
\includegraphics[width=0.9\textwidth]{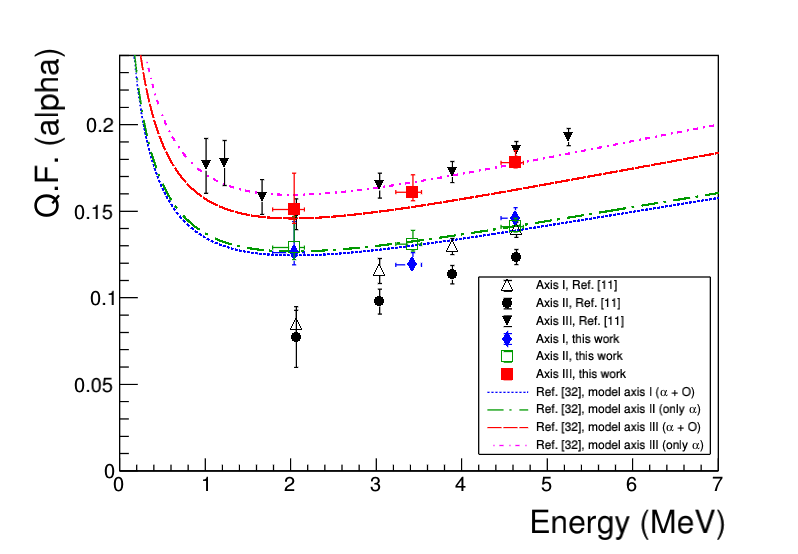}
\caption{\footnotesize{Dependence of the Q.F. for $\alpha$ particles on the energy measured with a ZnWO$_4$ 
scintillator
in Ref. \cite{Danev05} (black points) compared with those performed in this work (colored points). 
The crystal was irradiated in the directions along the crystal axes I (blue on-line), II (green on-line), and
III (red on-line), respectively. 
The anisotropic behaviour of the crystal is evident.
The models for each crystal axis obtained as global (or only $\alpha$) fits on the data of this paper
following the prescription of Ref. \cite{Tret2010} are also reported. See text.
}}
\label{fg:q_alpha}
\end{figure}

As shown in Fig. \ref{fg:q_alpha}, the quenching factor for $\alpha$ particles
measured along the crystal axis III is about 20\% higher than
that measured along the crystal axes I and II. 
Instead, the quenching factor measured along the crystal axes I and II are quite similar. The error bars are 
mainly due to the uncertainty of the alpha energy slightly degraded in air.

The quenching factors and the anisotropic effect measured here are in reasonable agreement with those reported in Ref. \cite{Danev05}, 
as shown in Fig. \ref{fg:q_alpha}. 

Fig. \ref{fg:q_alpha} also shows the models for each crystal axis obtained as global fit on the data
of this paper following the prescription of Ref. \cite{Tret2010},
based on the semi-empirical classical Birks formula \cite{Birks94}:

\begin{equation}
\frac{dL}{dr} =  \frac{S\frac{dE}{dr}}{1+kB\frac{dE}{dr}},
\end{equation}

\noindent where $S$ is a normalization factor, not relevant here, 
$BdE/dr$ is the density of excitation centers along the track, and $k$ is a parameter; 
$kB$ is usually treated as a single parameter (Birks factor).
As done in Ref. \cite{Tret2010}, a single $kB$ parameter has been used to fit the data
of $\alpha$ particles and recoils (see Section \ref{meas_qf}).
Moreover, the fits of the only $\alpha$ data for axes II and III are reported in Fig.  \ref{fg:q_alpha} for further discussion.

In conclusion, the data confirm the anisotropic features of ZnWO$_4$ crystal scintillator in the case of a beam of $\alpha$ particles.

\section{Measurements with neutrons}
\label{meas_n}

In order to verify and quantify scintillation anisotropy for low energy recoiling nuclei,
the quenching factors
for each crystallographic axis of the ZnWO$_4$ crystal have been measured with the set-up 
described in the following.
In particular, the nuclear recoils are induced by neutrons produced by a neutron generator;
after a single elastic scattering, the neutron is tagged by two neutron detectors placed at a given scattering angle. 
This defines the energy and the direction of the recoiling nucleus, while the detected energy in the ZnWO$_4$ detector
-- measured in keV electron equivalent (keVee) -- allows the determination of the quenching factor, depending on the 
relative position of the crystal axes.

\subsection{The neutron generator}

The neutrons were generated by means of a neutron generator. 
In particular, a portable generator from 
Thermo Scientific (Model MP 320) has been used. In this device, deuterons
are accelerated in electric potential up to 80 kV 
with available beam currents in the range 20-70 $\mu$A. This 
particular device is endowed with a tritium target 
so that neutrons produced in the $d(t,\alpha)n$ reaction have energies around 14.7 
MeV in the forward direction. The experimental 
beam parameters were chosen in order to maximize the neutron yield 
at a stable beam operation. 
The $d(t,\alpha)n$ cross section peaks at a deuteron beam energy of 160 keV, which is
well beyond the attainable range of the generator. However, considering the power-law 
behaviour of the cross section as a function of the 
incident energy, no serious loss is suffered (roughly speaking, around 
20\%) in the neutron yield by choosing a beam 
acceleration voltage of 60 kV with a beam current I=40 $\mu$A in 
continuous operation. With these conditions, extremely 
stable operation of the tube has been observed for long measuring times 
with a production rate Y=10$^7$ n/s, which we 
deem sufficient for the counting statistics of our experiment. In the 
experimental setup, neutrons leaving the target 
at 90 degrees to the forward direction have been used; at this angle, 
simple kinematics predicts a value $E_n=14.05$ MeV 
for a beam acceleration voltage of 60 kV.

\subsection{The scheme of the experimental set-up}

A schematic view of the experimental set-up is reported in Fig. 
\ref{fg:schema}.
The neutron generator is contained in a wooden/paraffin box of about 10 cm of thickness.
In correspondence with the tritium target inside the neutron generator 
there is a squared open channel (8 $\times$ 8) cm$^2$ where the produced neutrons 
are free to impinge on the ZnWO$_4$ detector. Other paraffin shield is 
placed
ahead the box containing the neutron generator (see Fig. \ref{fg:photo1}). 
In such a way, the total length of the neutron channel is about 50 cm.

\begin{figure}[!th]
	\begin{center}
		\includegraphics [width=0.75\textwidth]{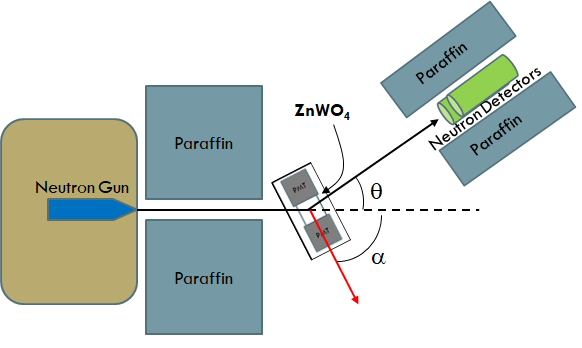}
	\end{center}
	\vspace{-0.8cm}
	\caption{\footnotesize{
		Schematic view from the top of the experimental set-up. The two neutron detectors are one above the other. 
	}}
	\label{fg:schema}   
\end{figure}
\normalsize

\begin{figure}[!th]
	\begin{center}
		\includegraphics [width=0.8\textwidth]{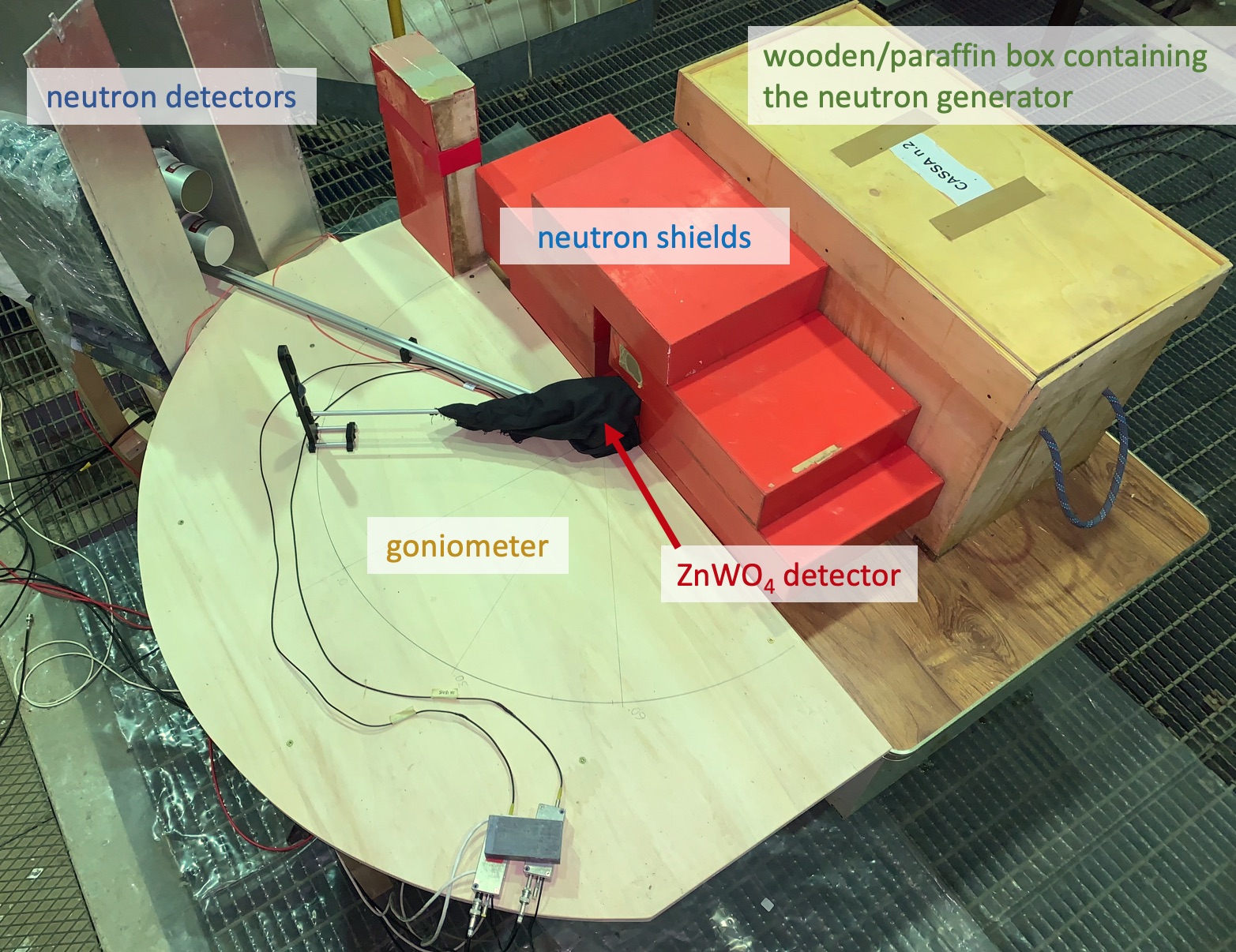}
	\end{center}
	\vspace{-0.8cm}
	\caption{\footnotesize{
		Picture of the experimental set-up during the installation. The wooden/ paraffin box can be seen in the background.
                The red paraffin bricks are used to further shield the neutron generator; they are around the neutron channel.
                The ZnWO$_4$ crystal coupled to two PMTs is inside a black plastic box; all is covered by a black rag.
	}}
	\label{fg:photo1}   
\end{figure}
\normalsize

The ZnWO$_4$ detector is placed in front of the neutron channel on a 
revolving platform around a vertical axis ($O$ axis),
which allows us to fix the direction of the crystal axis with respect to 
the impinging neutrons: the $\alpha$ angle as depicted in Fig. \ref{fg:schema}.
The ZnWO$_4$ crystal is centered on the $O$ axis and is optically coupled to two 
HAMAMATSU H11934-200 PMTs on opposite faces.
In the set-up, the PMTs are on the faces perpendicular to the crystal axis 
III; thus, $\alpha$ identifies the angle between the
crystal axis III and the impinging neutrons direction. The crystal axis I 
is also on the horizontal plane, while 
the axis II is vertical. The ZnWO$_4$ crystal and the two PMTs are inside a 
black plastic box, as shown in the picture on
Fig. \ref{fg:photo1}.

The neutrons scattered off the ZnWO$_4$ crystal are tagged by two neutron 
detectors by Scionix \cite{scionix}; they have an active volume: 76 mm diameter $\times$ 76 mm height, 
filled with EJ-309 liquid scintillator and a 
read-out by a 76 mm diameter ET 9821 PMT. 
The EJ-309 scintillator has a very high 
capability to discriminate neutrons interactions 
from the gamma background by Pulse Shape Discrimination (PSD). 

The neutron detectors are placed 82 cm far from $O$ axis, one above the 
other (see Fig. \ref{fg:photo1}), 
to maintain the same neutron scattering angle $\theta$ (see Fig. 
\ref{fg:schema}) and to improve the solid angle acceptance.
The distance center-to-center is 13 cm and the center of the bottom 
(upper) EJ-309 detector is 13 (117) mm under (above) 
the horizontal plane crossing the center of the ZnWO$_4$ detector.
They are held by an arm, shown in Fig. \ref{fg:photo1}, free to rotate 
around the $O$ axis.
It is useful to define a third angle, $\alpha_R$, for each different recoiling nucleus:
it is the angle between the nuclear recoil direction and the impinging 
neutron direction,
obtained by the kinematics of the elastic scattering of neutrons off nuclei 
once the $\theta$ angle is fixed.

\begin{table}[!hbt]
	\caption{\footnotesize{Kinematic variables in the case under study. The energy 
of the impinging neutrons is 14.05 MeV. 
For each scattering angles, $\theta$, and for the three nuclei of the ZnWO$_4$ crystal, 
it is reported the energy, E$_R$, and the angle, $\alpha_R$, of the recoiling nucleus, 
the energy, E$'_n$, and the velocity, v$'_n$, of the scattered neutrons.
The Time of Flight ($TOF$), that is the time between the elastic scattering on the ZnWO$_4$ crystal 
and the interaction on the neutron EJ-309 detectors, 82 cm far away, is around 16 ns in all cases.}}
	\begin{center}
		\resizebox{0.7\textwidth}{!}{ 
			\begin{tabular}{c|c|r|c|c|c}\hline
			$\theta$      & Recoiling   & \multicolumn{1}{c|}{E$_R$} & $\alpha_R$     &  E$'_n$  &  v$'_n$             \\
   			                   & nucleus     & \multicolumn{1}{c|}{(keV)} &                &  (MeV)   &  (m/s)                 \\		
			\hline\hline
                                      & Zn          &  216.7    & 59.6$^{\circ}$ & 13.835   &  $5.09\times 10^7$        \\
         		60$^{\circ}$  & W           &   77.5    & 59.9$^{\circ}$ & 13.975   &  $5.11\times 10^7$        \\
                                      & O           &  865.7    & 58.4$^{\circ}$ & 13.186   &  $4.97\times 10^7$         \\
			\hline
                                      & Zn          &  284.5    & 54.6$^{\circ}$ & 13.768   &  $5.08\times 10^7$       \\
         		70$^{\circ}$  & W           &  101.9    & 54.9$^{\circ}$ & 13.950   &  $5.11\times 10^7$        \\
                                      & O           & 1128.2    & 53.3$^{\circ}$ & 12.924   &  $4.92\times 10^7$        \\
			\hline
                                      & Zn          &  356.4    & 49.6$^{\circ}$ & 13.696   &  $5.06\times 10^7$     \\
         		80$^{\circ}$  & W           &  127.8    & 49.8$^{\circ}$ & 13.924   &  $5.11\times 10^7$        \\
                                      & O           & 1402.1    & 48.2$^{\circ}$ & 12.650   &  $4.87\times 10^7$        \\
			\hline
			\end{tabular}
		}
		\label{tb:kinematics}
	\end{center}
\end{table}

Some examples of the kinematic variables at different scattering angles 
and for the three nuclei of the ZnWO$_4$ crystal are reported in Table 
\ref{tb:kinematics}.
There is also reported the recoil energy, E$_R$, 
the energy, E$'_n$, and the velocity, v$'_n$, of the scattered neutrons.
The Time of Flight ($TOF$),
that is the time between the elastic scattering on the ZnWO$_4$ crystal 
and the interaction on one of the two neutron EJ-309 detectors,
82 cm far away, is around 16 ns in all cases;
the associated uncertainty has been evaluated by a Monte Carlo simulation as
0.5 ns ($RMS$), mainly due to the geometrical shape of the neutron detectors.
It is also worth noting that Table \ref{tb:kinematics} also includes W and Zn recoiling nuclei, 
even if the experiment is not sensitive to them, as described in the following.

\subsection{The electronic chain}

The data acquisition is based on the use of a CAEN DT5720 transient 
digitizer with 250 MSamples/s, 12 bit, 4-channel.
The signals from the two PMTs coupled to the ZnWO$_4$ detector are summed; a sum signal
is fed to the channel 0 of the digitizer.
\begin{figure}[!th]
	\begin{center}
		\includegraphics [width=0.9\textwidth]{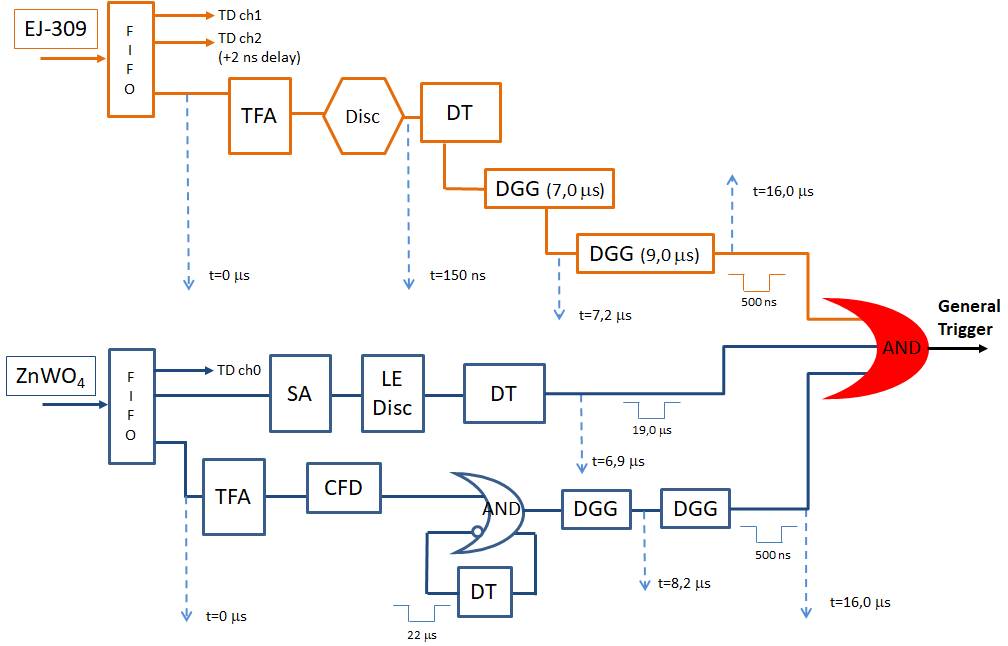}
	\end{center}
	\vspace{-0.8cm}
	\caption{\footnotesize{
		Simplified electronic scheme. The upper part describes the trigger generation of the EJ-309 detectors,
                while the bottom part represents the trigger chain for the ZnWO$_4$ detector. The triple coincidence (red on-line)
                between the EJ-309, the presence of energy-above-threshold in ZnWO$_4$, and the fast signal in ZnWO$_4$ 
                provides the general trigger to the data acquisition, as described in the text. TFA is The Time Filter Amplifier, Disc is a discriminator,
                CFD is a constant fraction discriminator, DT and DGG are delay gate generators, and SA is the spectroscopy amplifier.
	}}
	\label{fg:elect}   
\end{figure}

The signals from the two EJ-309 neutron detectors are also summed and two 
copies are fed to the channels 1 and 2 of the digitizer.
The first one is directly connected to the digitizer, the other one has a 
2 ns delay; this permits to obtain interleaving 
sampling to increase the sampling frequency because of the fast decay time 
of the EJ-309 scintillator.

The trigger is obtained by the coincidence between a signal in the 
ZnWO$_4$ detector and in the EJ-309 detectors
within $\pm 500$ ns. The simplified electronic scheme is shown in Fig. 
\ref{fg:elect}.
The upper part of the electronic scheme describes the trigger generation of the 
EJ-309 detectors,
while the bottom part represents the trigger chain for the ZnWO$_4$ 
detector. 
In particular, a copy of the ZnWO$_4$ signal is shaped and amplified by a 
Spectroscopy Amplifier ORTEC 672, and discriminated to provide
a energy-above-threshold logic signal. Another copy is amplified by a TFA 
ORTEC 474, which reduces the bandwidth, and discriminated
for having a fast time trigger. The general trigger is issued about 16 
$\mu$s after the original pulses. This allows the recording of the
tracks by the digitizer over a time window of 50 $\mu$s, with 20 $\mu$s of pre-trigger
(this implies that the digitized tracks have at least the 4 $\mu$s before
the occurrence of the signal).

\subsection{The data taking}

The energy detected in the ZnWO$_4$ detector is reconstructed by calculating the digitized pulse area.
The energy scale was calibrated with $^{133}$Ba and $^{137}$Cs sources.
The typical energy resolution was $\sigma/E = 4.4\%$ at the 662 keV peak.
The peak position was monitored before and after the data collection with neutrons; each data run lasts about two hours.
To monitor the stability of the energy scale, the data have been grouped in several runs and the stability in time of the energy spectra
has been verified. Moreover, the set-up was positioned inside a shelter where the temperature was maintained at reasonably constant level. 
However, the typical light output variation in ZnWO$_4$ detectors is around $-1.4\%/^\circ$C at room temperature (see Ref. \cite{direz3} and references therein). 
Thus, even a $2 ^\circ$C temperature variation implies 3\% in the light yield of the crystal. 
Therefore, any drift during the data collection can be safely excluded.
Fig. \ref{fg:cs137} shows a typical energy distribution of the ZnWO$_4$ detector for $\gamma$'s from $^{137}$Cs source.

\begin{figure}[!th]
	\begin{center}
		\includegraphics [width=0.6\textwidth]{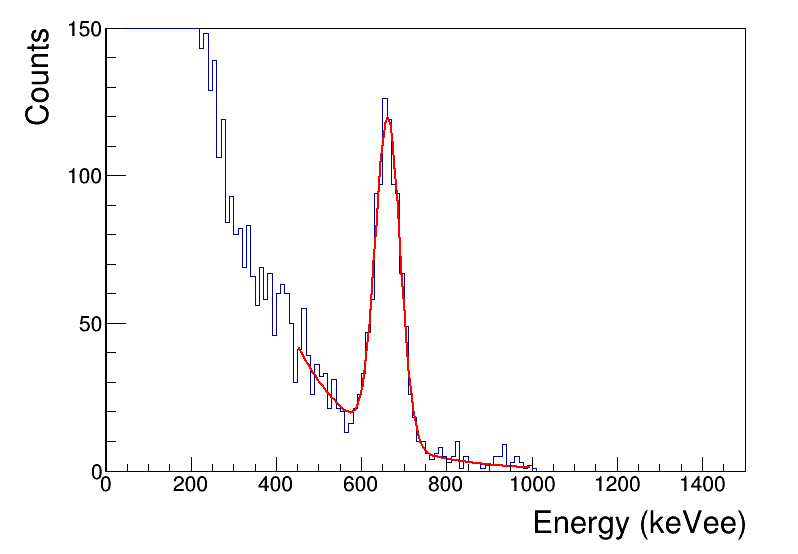}
	\end{center}
	\vspace{-0.8cm}
	\caption{\footnotesize{
		A typical energy distribution of the ZnWO$_4$ detector for $\gamma$'s from $^{137}$Cs source. A gaussian fit of the
                662 keV peak is shown.
	}}
	\label{fg:cs137}   
\end{figure}

Although the hardware energy threshold of the ZnWO$_4$ detector is maintained as low as possible accordingly to the electronic noise, 
a software energy threshold of 30 keVee is applied to the off--line analysis of the neutron data in order to
cut the large number of events due to inelastic processes, activation and background.
Thus, the oxygen recoils can be properly observed, while 
the Zn and W recoils are expected to be at lower energies (between 7 keVee and 14 keVee for Zn recoils, and
between 1.6 keVee and 2.5 keVee, for W recoils in the measurements here reported),
considering the expected quenching factors using the extrapolation of Ref. \cite{Tret2010}, and
the kinematical energies (see Table \ref{tb:kinematics}). 

Concerning the neutron detectors EJ-309, two different techniques are used to extract the information on
the PSD: i) head/tail analysis; ii) the analysis of the mean time, $\tau$, of the time profile of 
the EJ-309 pulse. In particular, the head is defined as the area of the pulse between the starting time
of the pulse, $t_0^{EJ}$, up to $t_0^{EJ}+40$ ns; the tail is defined as the area between $t_0^{EJ}+40$ ns and $t_0^{EJ}+50$ ns.
The $\tau$ variable is defined as: 

$$\tau= \frac{\Sigma_ih_it_i}{\Sigma_ih_i},$$

\noindent where $h_i$ is the pulse height at the time $t_i$ beginning from the starting time of the signal; the sum is extended over the first 100 ns of the pulse.
Typical examples of the separation between gamma and neutrons in both cases are reported in Fig. \ref{fg:psd}.

\begin{figure}[!th]
	\begin{center}
		\includegraphics [width=0.49\textwidth]{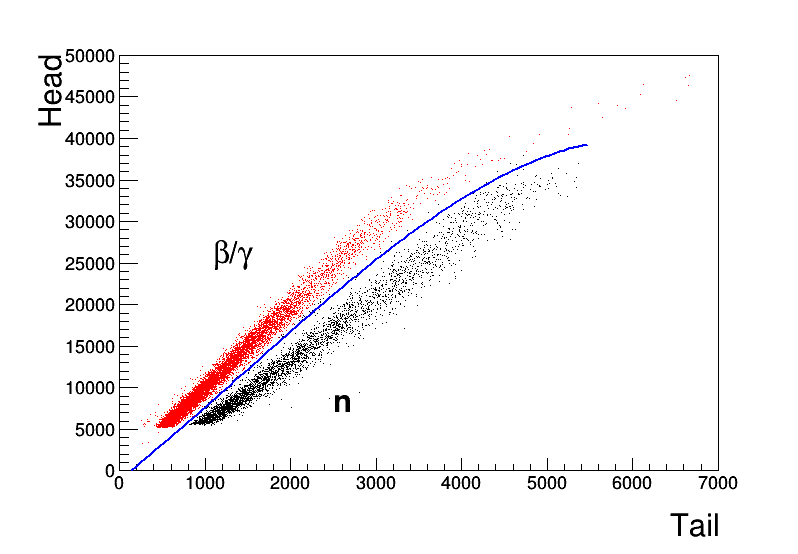}
		\includegraphics [width=0.49\textwidth]{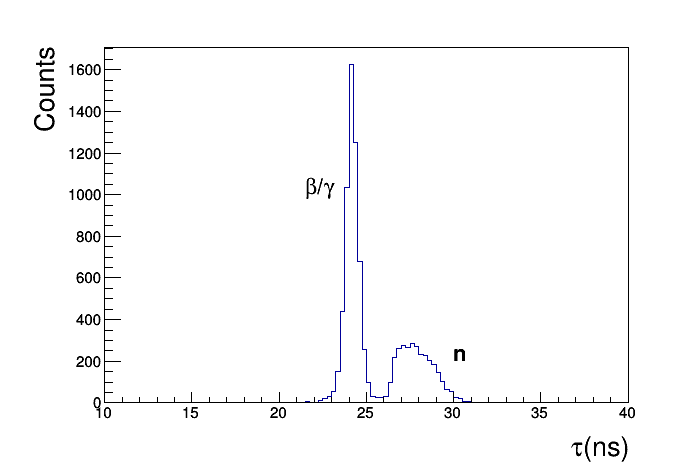}
	\end{center}
	\vspace{-0.8cm}
	\caption{\footnotesize{
		Example of gamma/neutron separation by PSD in EJ-309 liquid scintillator. Left: head/tail 
                analysis. Right: distribution of the mean time, $\tau$, variable. 
	}}
	\label{fg:psd}   
\end{figure}

An important tool to select the events related to the searched effect (elastic scattering of neutrons in the ZnWO$_4$ crystal) over 
the background is the study of the $TOF$ between the ZnWO$_4$ detector
and the neutron detectors EJ-309.
For this purpose we define the variable $\Delta t = t_0^{EJ} - t_0^{ZWO}$, where $t_0^{EJ}$ ($t_0^{ZWO}$)
is the starting time of the EJ-309 (ZnWO$_4$) pulse.
Since the transit time of the used PMTs is different (in particular, it is $\approx 40$ ns for the ET 9821 \cite{pmt_et}
and $\approx 6$ ns for the HAMAMATSU H11934-200 \cite{pmt_hama}), the $\Delta t$ variable is shifted with respect to 
$TOF$ and must be properly calibrated.
For this purpose we select coincidences between high energy events in the ZnWO$_4$ detector
and gamma events in the neutron detectors.
These coincidences are mostly due to either $(n,\gamma)$ interactions or Compton $\gamma$ scatterings, from the neutron generator.
The secondary gamma rays are observed in the neutron detectors; we select high energy events in ZnWO$_4$ for having a clean sample of coincidences 
with a well-defined starting time of the ZnWO$_4$ electronic pulse. 

The corresponding $\Delta t$ distribution is reported in Fig. \ref{fg:tof}.
A clear peak is present around $\Delta t \approx 34$ ns (where it is expected); actually, this is consistent with $TOF=2.7$ ns, 
the time required for a $\gamma$ quantum to travel from the ZnWO$_4$ detector to the neutron detectors. 
This calibration of the $TOF$ is applied in the following analysis.

\begin{figure}[!th]
	\begin{center}
		\includegraphics [width=0.6\textwidth]{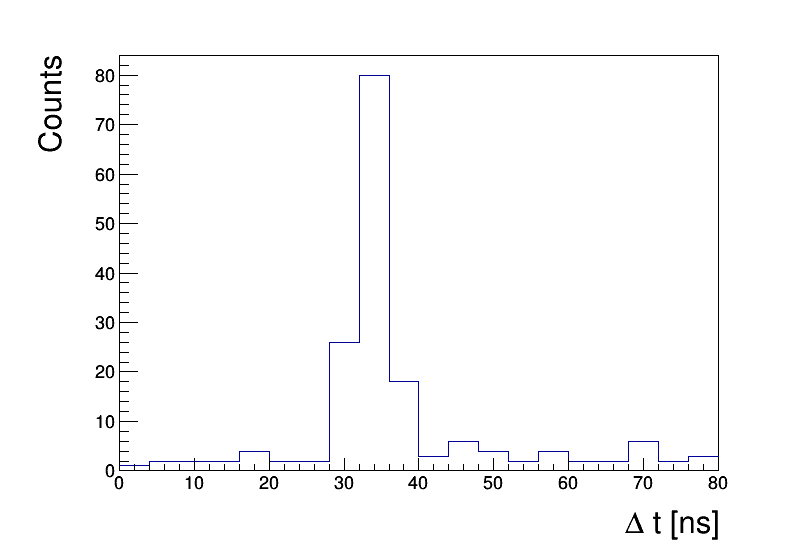}
	\end{center}
	\vspace{-0.8cm}
	\caption{\footnotesize{
	  $\Delta t$ distribution for coincidences between high energy events in the ZnWO$_4$ detector
and gamma events in the neutron detectors. 
A clear peak is present around $\Delta t \approx 34$ ns; actually, this is consistent with $TOF=2.7$ ns, the time required for a $\gamma$ quantum to travel from 
the ZnWO$_4$ detector to the neutron detectors. We recall that the sampling step of the digitizer is 4 ns.
	}}
	\label{fg:tof}   
\end{figure}

\subsection{The measured quenching factors}
\label{meas_qf}

The data collected during the neutron irradiation at fixed scattering angle $\theta$ and crystal axis\footnote{The typical
live time for each data set ranges from 11 to 25 h depending on the scattering angle and crystal axis.} 
have been analysed 
by selecting coincidences between events in the ZnWO$_4$ detector and neutrons in the neutron detectors.
An example of the $TOF$ distribution and of the bi-dimensional plot $TOF$ vs ZnWO$_4$ energy ($E_{ZWO}$, in keVee) 
is reported in Fig. \ref{fg:tof_n}
for the case of scattering angle $\theta=70^{\circ}$ and ZnWO$_4$ crystal axis I.
These plots show a continuum due to random coincidences and a clear peak 
which is well in agreement with the expected $TOF$ for neutrons after elastic 
scattering off ZnWO$_4$ detector (see Table \ref{tb:kinematics}). The peak in the $TOF$ variable shows a tail on the left
due to the first photoelectron delay in ZnWO$_4$ (effective average scintillation decay time $\approx$ 24 $\mu$s \cite{Danev05}).
Thus, the mean value of the $TOF$ variable for neutrons after elastic 
scattering off ZnWO$_4$ detector is systematically smaller than the values reported in Table \ref{tb:kinematics}
owing to the prompt response of the neutrons detectors and to the possible delay of the first photoelectron in the ZnWO$_4$ scintillator.
The peak in the $E_{ZWO}$ variable is due to the oxygen recoils in the ZnWO$_4$ detector 
and its position provides the quenching factor of oxygen in ZnWO$_4$ for the used scattering angle and crystal axis.

\begin{figure}[!ht]
	\begin{center}
		\includegraphics [width=0.49\textwidth]{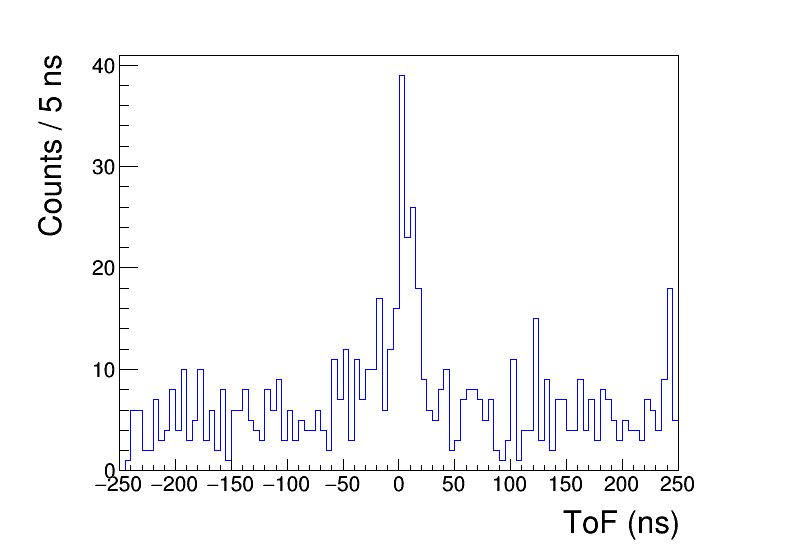}
		\includegraphics [width=0.49\textwidth]{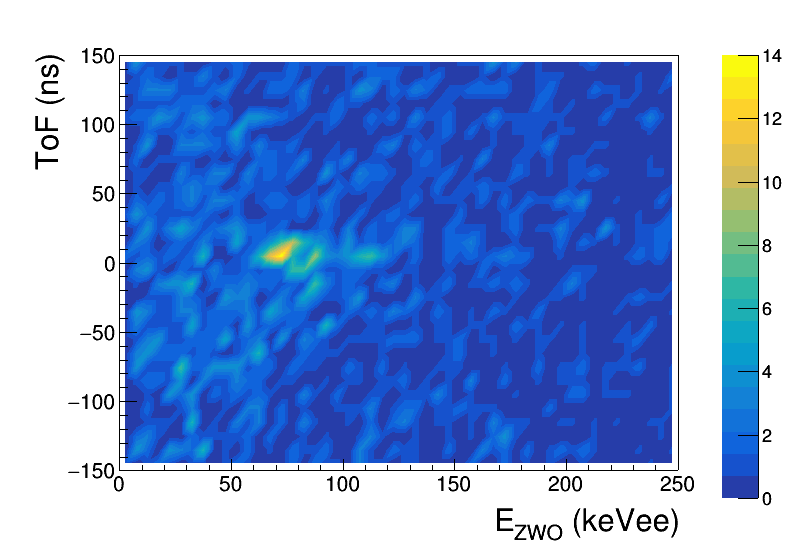}
	\end{center}
	\vspace{-0.8cm}
	\caption{\footnotesize{
                 Examples of the $TOF$ distribution and of the bi-dimensional plot $TOF$ vs ZnWO$_4$ energy (in keVee) 
                 for coincidences obtained after selecting neutrons in the neutron detectors,
                 for the case of scattering angle $\theta=70^{\circ}$ and axis I.
		 Left: The $TOF$ distribution shows a continuum due to random coincidences and a clear peak, 
                 that is in agreement with the expected $TOF$ for neutrons after elastic 
                 scattering off ZnWO$_4$ detector (see Table \ref{tb:kinematics}). The peak shows a tail on the left
                 due to the first photoelectron delay in ZnWO$_4$.
		 Right: A clear peak is present at the proper value of $TOF$ and at a given energy. The peak position 
                 in the $E_{ZWO}$ distribution
                 provides the quenching factor for the used scattering angle and crystal axis.
	}}
	\label{fg:tof_n}   
\end{figure}

\begin{figure}[!ht]
	\begin{center}
		\includegraphics [width=0.49\textwidth]{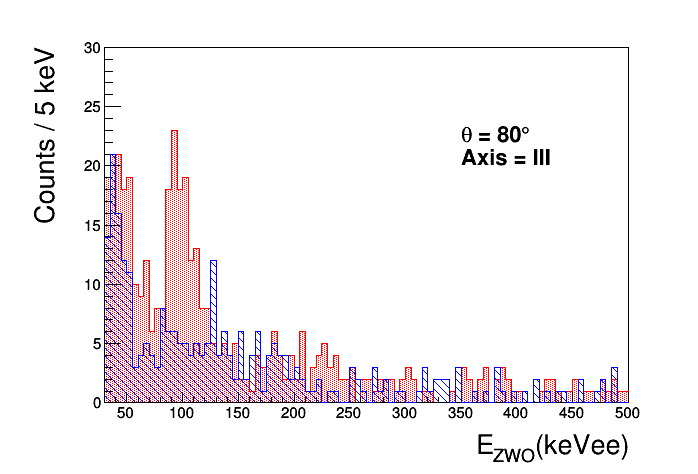}
		\includegraphics [width=0.49\textwidth]{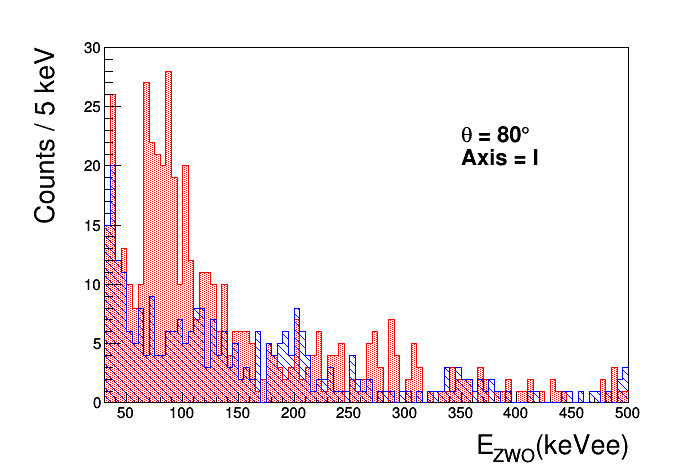}
		\includegraphics [width=0.49\textwidth]{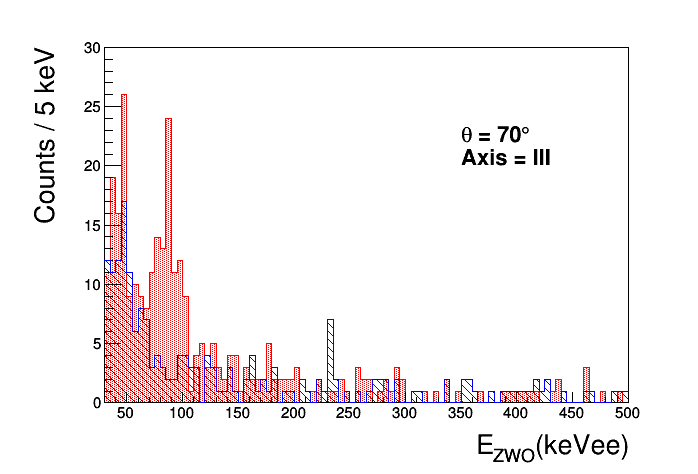}
		\includegraphics [width=0.49\textwidth]{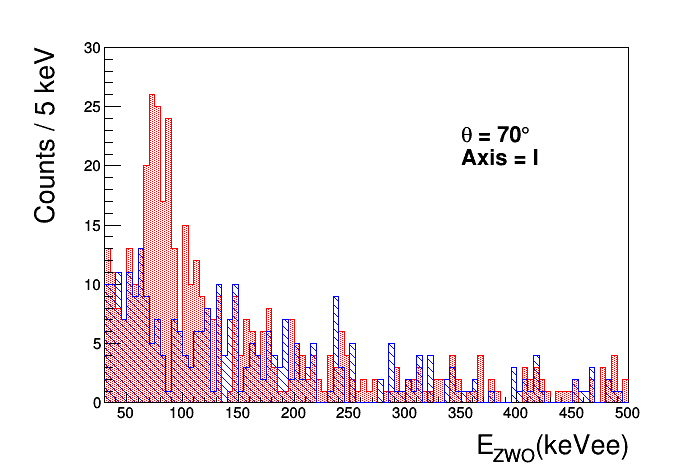}
		\includegraphics [width=0.49\textwidth]{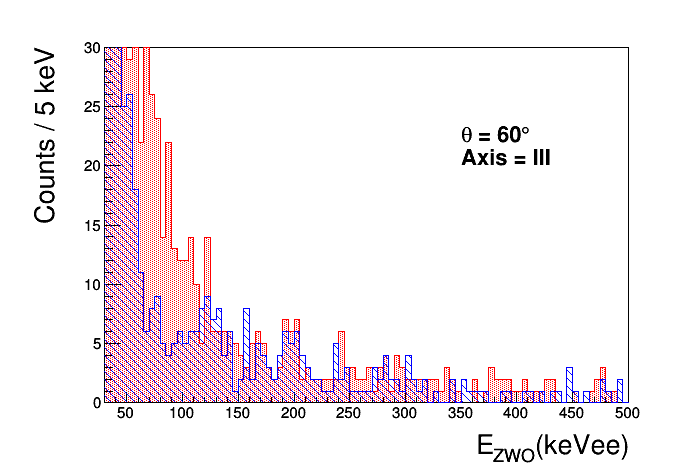}
		\includegraphics [width=0.49\textwidth]{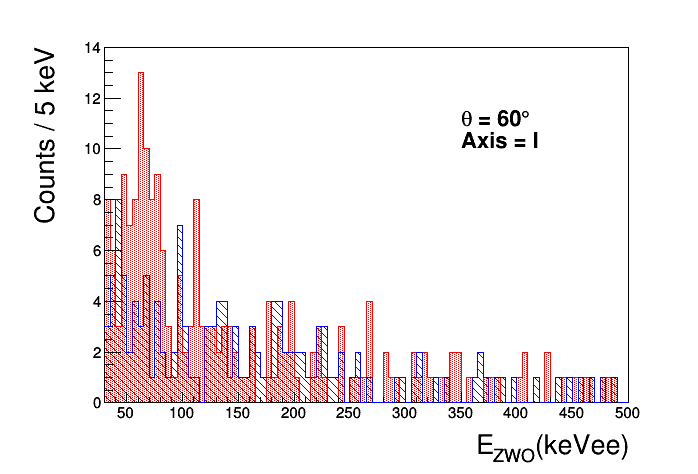}
	\end{center}
	\vspace{-0.8cm}
	\caption{\footnotesize{
Energy distributions of the detected energy, $E_{ZWO}$, in the ZnWO$_4$ detector, measured in keV electron equivalent (keVee)
for the six configurations (3 scattering angles and axes III and I). Only events identified as neutrons in the neutrons detectors are selected.
Light (red on-line) histogram: events selected in the proper window of the $TOF$ variable
($-20$ ns $< TOF < 30$ ns).
Dashed (blue on-line) histogram: events selected in the off-window ($60$ ns $< TOF < 110$ ns).
	}}
	\label{fg:ezwo}   
\end{figure}
 
Three scattering angles were considered. Since the responses of the crystal axes I and II are rather
similar, only the case of axis I was considered.
Fig. \ref{fg:ezwo} shows the energy
distributions of the detected energy, $E_{ZWO}$, in the ZnWO$_4$ detector, measured in keV electron equivalent, for the six configurations (3 scattering angles and axes III and I).
Two distributions are reported for each plot: one is obtained by selecting events in the time window 
expected for the $TOF$ of neutrons elastically scattered off ZnWO$_4$ detector,
$-20$ ns $< TOF < 30$ ns (light red histogram); the other is presented by selecting events in
off-window (random coincidences), $60$ ns $< TOF < 110$ ns (blue histogram).
The number of the selected events in the in-window (off-window) histograms 
in the six configurations runs from 211 to 792 (148 to 536).

The off-window events are related to random coincidences and, therefore, their distribution
is the background distribution in the histogram of the in-window events.
The peaks of oxygen nuclear recoils are rather evident.
The positions of these peaks are obtained by fitting them by a gaussian curve plus an exponential,
that simulates the background of the random coincidences.
For the only case of $\theta=60^{\circ}$ for axis III, where such a method is less accurate,
we adopt a slight different procedure: the spectrum obtained by subtracting the
background of the off-window events has been fitted by a gaussian.
The energy resolution, $\sigma$, of these peaks are between 8 and 12 keVee, values well in agreement
with the hypothesis that the dependence of the energy resolution on the energy 
can be described by a root function $1/\sqrt E$.

The peak positions
are reported in Table \ref{tb:scenarios}, where the expected recoil energies for the oxygen nucleus, $E_{R,O}$, are also noted.
The latter ones are verified by a Monte Carlo simulation including the kinematics of the process (see also Table 
\ref{tb:kinematics});
the intrinsic spread of the expected recoil energies are then evaluated to be of the order of 2-3 keVee,
well below the energy resolution of the detector.
The obtained quenching factors are calculated as the ratio $E_{ZWO}/E_{R,O}$ and the last column of Table \ref{tb:scenarios}
reports the degree of anisotropy for each scattering-angle/recoil energy.
Finally, we can define a ``mean anisotropy'' as the average of the three $Q_{III}/Q_{I}$ values. It is: $(1.145 \pm 0.027)$, 
giving an evidence of anisotropy in the light response for oxygen nuclear recoils at 5.4 standard deviations confidence level.

\begin{table}[!ht]
	\caption{\footnotesize{Summary table of the peak position due to oxygen nuclear recoils.
For each scattering angles, $\theta$, and for the different axes of the ZnWO$_4$ crystal, 
it is reported the peak position, $E_{ZWO}$, the energy resolution, $\sigma$,
and the expected recoil energies for the oxygen nucleus, $E_{R,O}$.
The quenching factors, $Q$, and the anisotropy, that is the $Q_{III}/Q_{I}$ ratio, are reported 
in the last columns.}}
	\begin{center}
		\resizebox{0.9\textwidth}{!}{ 
			\begin{tabular}{c|c|c|c|r|c|c}\hline
			Scattering angle, & Crystal  & $E_{ZWO}$       &  $\sigma$  &    \multicolumn{1}{c|}{$E_{R,O}$}       &  Quenching            &  $Q_{III}/Q_{I}$                   \\
			$\theta$               & axis       & (keVee)              &  (keVee)   &    \multicolumn{1}{c|}{(keV)}           &  factor, $Q$          &                                    \\		
			\hline\hline
    \multirow{2}{*}{80$^{\circ}$}     & III      &  $99.3\pm2.5$   & \,\;9    & \multirow{2}{*}{1402}    &  $0.0708\pm0.0018$    & \multirow{2}{*}{$1.174\pm0.051$}  \\
			                           & I        &  $84.5\pm2.9$   &   12    &                                      &  $0.0603\pm0.0021$    &   \\
			                           &    &  & &     &  &   \\
    \multirow{2}{*}{70$^{\circ}$}     & III      &  $86.5\pm2.0$   & \,\;7    & \multirow{2}{*}{1128}    &  $0.0767\pm0.0018$    & \multirow{2}{*}{$1.121\pm0.038$}  \\
			                           & I        &  $77.2\pm1.9$   &   10    &                                      &  $0.0684\pm0.0017$    &   \\
			                           &    &  & &     &  &   \\
    \multirow{2}{*}{60$^{\circ}$}     & III      &  $75.4\pm1.8$   & \,\;9    & \multirow{2}{*}{866}      &  $0.0871\pm0.0021$    & \multirow{2}{*}{$1.166\pm0.059$}   \\
			                           & I        &  $64.7\pm2.9$   &   10    &                                      &  $0.0747\pm0.0033$    &   \\
			\hline
			\end{tabular}
		}
		\label{tb:scenarios}
	\end{center}
\end{table}

\begin{figure}[!h]
	\begin{center}
		\includegraphics [width=0.8\textwidth]{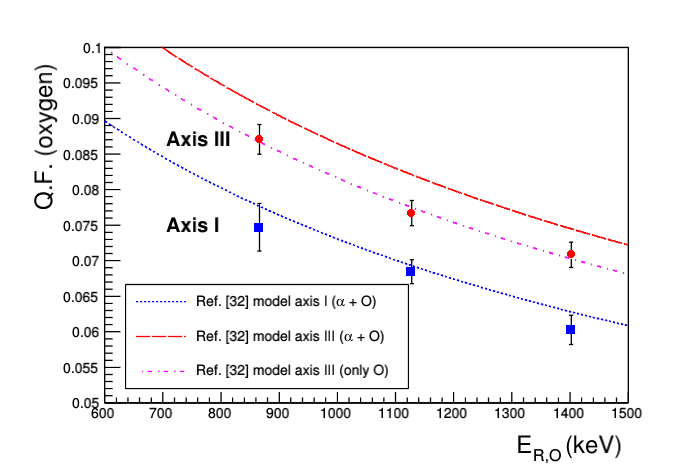}
	\end{center}
	\vspace{-0.8cm}
	\caption{\footnotesize{
Quenching factors for oxygen nuclear recoils in ZnWO$_4$ for the crystal axes I and III as function of the expected recoil energies $E_{R,O}$.
The models for the considered crystal axes obtained as global fits on the data of this study ($\alpha$ and oxygen recoil data)
following the prescription of Ref. \cite{Tret2010} are also reported. 
It is also shown the fit of the only oxygen recoil data points for axis III. See text.
	}}
	\label{fg:qf}   
\end{figure}

The obtained quenching factors are also depicted in Fig. \ref{fg:qf}.
There are also shown the models for the considered crystal axes obtained as global fits on the data of this paper ($\alpha$ and oxygen recoil data, ``$\alpha +$O'')
following the prescription of Ref. \cite{Tret2010}. A single $kB$ parameter has been used to fit the data
of $\alpha$ particles and oxygen recoils for each axis. They are 
$kB = 12.06$ mg MeV$^{-1}$ cm$^{-2}$ for axis I,
$kB = 11.80$ mg MeV$^{-1}$ cm$^{-2}$ for axis II (only $\alpha$ data are considered, ``only $\alpha$'' in Fig. \ref{fg:q_alpha}), and
$kB = 9.98$ mg MeV$^{-1}$ cm$^{-2}$ for axis III, respectively.

The global fit for axis I shown in Figs. \ref{fg:q_alpha} and \ref{fg:qf} is rather good;
for axis III the trend is qualitatively well represented, anyhow a single $kB$ value appears not sufficient to exactly reproduce both data sets:
$\alpha$ and oxygen recoils (``$\alpha +$O'').
Thus, only for axis III the fits have also been done separately, giving good agreements with the data. The obtained $kB$ values are:
$kB = (8.95^{+0.19}_{-0.26})$ mg MeV$^{-1}$ cm$^{-2}$ for $\alpha$ data points (``only $\alpha$'' in Fig. \ref{fg:q_alpha});
$kB = (10.65\pm0.17)$ mg MeV$^{-1}$ cm$^{-2}$ for oxygen recoil data points (``only O'' in Fig. \ref{fg:qf}).

For completeness, we also recall the results of Ref. \cite{Biz2012}, where the quenching factors for ions (including oxygen) in CdWO$_4$ 
crystal scintillator were measured with the help of an accelerator.  The CdWO$_4$ quenching factors for the crystal plane (001) were given
in Fig. 1 of Ref. \cite{Biz2012}. The ZnWO$_4$ and CdWO$_4$ scintillators have rather similar properties, also including the presence of 
anisotropic properties.

\section{Conclusions}

The data presented here confirm the anisotropic response of the ZnWO$_4$ crystal scintillator to $\alpha$ particles in the MeV energy
region. The anisotropy is significantly evident also 
for oxygen nuclear recoils in the energy region down to some hundreds keV at 5.4 $\sigma$ confidence level.
Moreover, the presence of a good anisotropic response also in lower energy region
is supported by the trend of the measured quenching factors with lower energy,
according to which, as shown in Fig. \ref{fg:qf}, the quenching factors of nuclear recoils improve going to lower energy.

\section{Acknowledgement}

It is a pleasure to thank Mr.~A.~Bussolotti and Mr.~A.~Mattei for their valuable technical support and Dr. M. Laubenstein for providing 
the silicon detector and for fruitful discussions.
The group from the Institute for Nuclear Research was supported in part by the program of the National Academy of Sciences of Ukraine ``Fundamental research on high-energy physics and nuclear physics (international cooperation)'' (Grant No. 0118U005411). D.V.K. and O.G.P. were supported in part by the project ``Investigations of rare nuclear processes'' of the program of the National Academy of Sciences of Ukraine ``Laboratory of young scientists'' (Grant No. 0118U002328).

\end{document}